\documentstyle[12pt, fleqn]{article}
\makeatletter
 
          \@addtoreset{equation}{section}
       \makeatother

\newcommand{\ymh}{Yang-Mills-Higgs Lagrangian$\;$}
\newcommand{\p}{\partial}
\newcommand{\nonum}{\nonumber\\}
\newcommand{\be}{\begin{equation}}
\newcommand{\ee}{\end{equation}}
\newcommand{\ba}{\begin{eqnarray}}
\newcommand{\ea}{\end{eqnarray}}
\newcommand{\bd}{\begin{displaymath}}
\newcommand{\ed}{\end{displaymath}}

\renewcommand{\a}{&\!\!\!\!\!&\!\!\!\!\!}
\newcommand{\ma}[1]{_{\mathop{}_{#1}}}
\begin{document}
\thispagestyle{empty}
\begin{center}
\vspace*{1.0cm}
{\Large \bf New Incorporation of  the Strong Interaction
\vskip 2mm
in NCG and Standard Model}
\vskip 2.5cm
{\large  Yoshitaka {\sc Okumura}\footnote{
e-mail address: okum@isc.chubu.ac.jp}}
\\
\vspace{3ex}
{\it  Department of Natural Science,
  Chubu University, Kasugai, 487, Japan}\\
    \vspace{2ex}
\vspace{2ex}
\vskip 2mm
\end{center}
\vspace{2.5 cm}
\begin{abstract}
The standard model is reconstructed by new method to incorporate
strong interaction into our previous scheme based
on the non-commutative geometry. The generation mixing is also
taken into account. Our characteristic point is to take the fermion
field so as to contain quarks and leptons all together which is almost
equal to that of SO(10) grand unified theory(GUT).
The space-time $M_4\times Z_2$;
Minkowski space multiplied by two point discrete space is prepared
to express the left-handed and right-handed fermion fields. The generalized
gauge field  $A(x,y)$ written in one-differential form extended on
$M_4\times Z_2$ is well built to give the correct Dirac
Lagrangian for fermion sector. The fermion field is a vector in
24-dimensional space and  gauge and Higgs fields are written in
$24\times24$ matrices.
At the energy of the equal coupling
constants for both sheets $y=\pm$  expected to be amount to the energy
of GUT scale, we can get $\sin^2\theta_{_{W}}=3/8$
and $m_{_{H}}=\sqrt{2}m_{_{W}}$. In general, the equation
$m\ma{H}=(4/\sqrt {3})m\ma{W}\sin\theta\ma{W}$ is followed.
Then, it should be noticed that the same result as that of
the grand unified theory such as  SU(5) or SO(10) GUT is obtained
without GUT but with the approach based on the non-commutative geometry
and in addition the Higgs mass is related to other physical quantities
as stated above.
\end{abstract}
\vfill\eject
\setlength{\baselineskip}{0.65cm}

\section{ Introduction}
Higgs mechanism is crucially important for all spontaneously broken gauge
theory such as the standard model, SU(5) or SO(10) grand unified theory.
Much efforts have been devoted to understand the essence of this mechanism.
However, not so much progress had been achieved so far before
the approach in non-commutative geometry(NCG) on the discrete space
$M_4\times Z_2$; Minkowski space multiplied by two points space
was proposed by Connes\cite{Con}. In this approach, Higgs field
due to this mechanism is regarded as a kind of the gauge field
on the discrete space, so that the origin of Higgs particle
becomes evident, and so any other extra physical modes are never needed.\par
Standard model which has cleared all experimental tests so far has been
reconstructed many times based on NCG from respective standpoints
\cite{Con}$\sim$\cite{KDC},\cite{Gou}.
We also reconstructed the standard model based on our formalism which
is the generalization of usual differential geometry on the
ordinary manifold. The extra differential one-form $\chi$
is introduced in addition to
the usual one-form $dx^\mu$ and so our formalism is very familiar
with the ordinary differential geometry, though original Connes'one is
very difficult to understand.
Our method is so flexible that it is  easily extended to reconstruct
the gauge theory with more complex structure such as the left-right
symmetric gauge theory, SU(5) or SO(10) GUT \cite{MO2},\cite{O10}.
We have made attempts several times
\cite{MO1},\cite{MO3},\cite{MO4},\cite{OSM}
to reconstruct the standard model
in which how to incorporate the strong interaction was different
with each other. The direct product method of gauge fields was
adopted in Ref.\cite{OSM} to take account of the strong interaction.
In addition, after the specifications of the fermion field on the
two point discrete space are determined
the generalized gauge field is defined to yield the correct Dirac
Lagrangian for fermion sector. This approach was suggested by Sogami
\cite{Soga} who started from the Dirac Lagrangian in the standard model
and defined the generalized covariant derivative
with gauge and Higgs fields operating on quark and lepton fields
to reproduce the correct Dirac Lagrangian.
The field strengths for both the gauge and Higgs fields are defined
by the commutators of the covariant derivative by which
he could obtain the Yang-Mills Higgs Lagrangian in the standard model
with the extra restriction on the coupling constant of the Higgs potential.
Recently, his method is developed to SU(5) GUT\cite{Soga2}.
\par
In this article, we also adopt this approach to reconstruct the standard model
and improve our previous article \cite{OSM} by taking the fermion field
$\psi(x,y)$ with the variable $y=\pm$ of two point discrete space
to be the fermion field of$\;$ SO(10) GUT with anti-fermion part discarded.
$\psi(x,y)$ is a vector in 24 dimensional space since
the three fermion families are also taken into account.
Then, we will decide the formations of gauge and Higgs fields which are
all expressed in $24\times 24$ matrices. It becomes possible to take the
left-handed gauge field, U(1) gauge field and color gauge field so as to
commute with each other.
According to our algebraic rules in NCG on the discrete space $M_4\times Z_2$,
we can get \ymh with the extra restriction on the coupling of Higgs
self interaction term within the assumption that the Sitarz term is neglected.
In the limit of the equal coupling
constants for both sheets $y=\pm$ which is amount to the energy
of GUT scale, we can get $\sin^2\theta_{_{W}}=3/8$
and $m_{_{H}}=\sqrt{2}m_{_{W}}$. In general, the equation
$m\ma{H}=(4/\sqrt {3})m\ma{W}\sin\theta\ma{W}$ is hold.
Then, it should be noticed that the same result as that of
the grand unified theory such as  SU(5) or SO(10) GUT is obtained
without GUT but with the approach based on the non-commutative geometry
and in addition, the Higgs mass is related to other physical quantities
as stated above.
\par
This article is divided into four sections. The next section presents
the modifications of our previous formalism
based on the generalized differential calculus on
$M_4\times Z_2$ so as to incorporate the generation mixing mechanism
and color symmetry.
In this section  a geometrical
picture for the unification of the gauge and Higgs fields is realized,
which is the ultimate understanding in this field.
The third section is the application to the standard model
which leads to the quite different predictions for particle masses
from the Sogami'one\cite{Soga}. This is because the complex Yukawa coupling
constants written in matrix form in the generation space is
not contained in the generalized gauge field in our formalism.
The last section is devoted to concluding remarks.
Appendix includes some remarks of our algebraic rules in NCG.
\section{ Basic Settings}
In this section  we slightly change our previous formulation \cite{MO1}
to construct the gauge theory in NCG on the discrete space
in order to correspond with the incorporation of the strong interaction
in the standard model.
This theme has been already treated in Ref.\cite{OSM} with the formalism
in which the generalized gauge field has the direct product form of
the color and flavour gauge fields, and the fermion field
is taken for each sheet of the discrete space
to contain leptons and quarks  all together in one form.
The idea to take the fermion field in such a form is inspired by Sogami
\cite{Soga} which defines the generalized covariant derivative with
gauge and Higgs fields operating on quark and lepton fields.
In this article, we change the formation of leptons and quarks in
a fermion field to match with the fermion field of SO(10) GUT
though the anti-fermions are discarded. As a result, we have to
devise to construct the generalized gauge field in order to
give the correct Dirac Lagrangian for fermion sector, which is the
purpose of this section. \par
In Ref.\cite{OSM}, the matrix $K$ in Ref.\cite{Cham}
was introduced to explain the generation mixing between leptons and
quarks. However, $K$ has no necessity to be introduced if we regard
$c\ma{Y}$ in the covariant
spinor one-form is related to the Yukawa coupling constant expressed
in matrix form of generation space, and moreover $K$ spoils
the beautiful relation between the Higgs and charged gauge boson
 masses in the limit of equal coupling constants for both sheets.
 Thus, we exile in this article the generation mixing matrix $K$
which is indispensable in Ref.\cite{Cham} to obtain the meaningful
Higgs potential.\par
Let us start with the equation of the generalized gauge field ${\cal A}(x,y)$
written in one-form on the discrete space $M_4\times Z_2$.
We modify it in the original form \cite{MO1} to take account of the
strong interaction in such a way that
\be
      {\cal A}(x,y)=\sum_{i}a^\dagger_{i}(x,y){\bf d}a_i(x,y)+
      \sum_{j}b^\dagger_{j}(x,y){d}b_i(x,y),\label{2.1}
\ee
where $x$ and $y\,(=\pm)$ are the variables in the Minkowski space $M_4$
and in the discrete space $Z_2$, respectively.
$a_i(x,y)$ and $b_j(x,y)$ are the square-matrix-valued functions
and commute with each other.
$i$ and $j$ are  variables of the extra
internal space which we can not now identify what it is.
As stated later, $\sum_{i}a^\dagger_{i}(x,y){\bf d}a_i(x,y)$
and $\sum_{j}b^\dagger_{j}(x,y){d}b_i(x,y)$ correspond with
the flavor and color gauge fields, respectively.
These equations of gauge fields
are very similar to the effective gauge field
in Berry phase \cite{Ber}, which might
lead to the identification of this internal space.
Now, we simply regard $a_i(x,y)$ and $b_j(x,y)$ as the more fundamental
fields to construct gauge and Higgs fields though they have only
mathematical meaning. $a_i(x,y)$ and $b_j(x,y)$ never appear
in final stage.
${\bf d}$ in Equation (\ref{2.1}) is the generalized exterior
derivative defined as follows.
\ba
&&       {\bf d}a_{i}(x,y)=(d + d_\chi)a(x,y)
                                   =(d+d_{\chi})a(x,y), \nonum
&&     da_i(x,y) = \partial_\mu a_i(x,y)dx^\mu,\hskip 1cm \label{Add1}\\
&&   d_{\chi} a_i(x,y) =[-a_i(x,y)M(y) + M(y)a_i(x,-y)]\chi,
        \label{2.2}\\
&&      {d}b_j(x,y)= \p_\mu b_j(x,y)dx^\mu \label{2.3}\\
&&   d_{\chi} b_j(x,y) =0. \label{Add2}
\ea
Here
$dx^\mu$ is
ordinary one-form basis, taken to be dimensionless, in Minkowski space
$M_4$, and $\chi$
is the one-form basis, assumed to be also dimensionless,
in the discrete space $Z_2$.
We have introduced $x$-independent matrix $M(y)$
whose hermitian conjugation is given by $M(y)^\dagger=M(-y)$.
The matrix $M(y)$ turns out to determine the scale and pattern of
the spontaneous breakdown of the gauge symmetry. Thus,
Equation(\ref{Add2}) means that the color symmetry
of the strong interaction does not break spontaneously.\par
In order to find the explicit forms of gauge and Higgs fields
according to Equations (\ref{2.1}) and (\ref{2.2}), we need the following
important algebraic rule of non-commutative geometry.
\be
        f(x,y)\chi=\chi f(x,-y), \label{2.4}
\ee
where $f(x,y)$ can be the quantity defined on the discrete space such as
$a_i(x,y)$, gauge field, Higgs field or fermion fields.
It should be noticed that Equation (\ref{2.4}) never expresses
the relation between
the matrix elements of $f(x,+)$ and $f(x,-)$ but insures the product between
the fields expressed in  differential form on the discrete space,
which can be easily seen in the calculation of the wedge product
${\cal A}(x,y)\wedge {\cal A}(x,y)$.
Eq.(\ref{2.4}) realizes the non-commutativity of our algebra in the geometry
on the discrete space $M_4\times Z_2$. In Appendix we will explain the
meaning of Eq.(\ref{2.4}) and some other rules in our NCG scheme
in more detail.
Using Eq.(\ref{2.4}) and some other algebraic rules,
${\cal A}(x,y)$ is rewritten as
\be
 {\cal A}(x,y)=A_\mu(x,y)dx^\mu+\Phi(x,y)\chi+G_\mu(x)dx^\mu, \label{2.5}
\ee
where
\ba
&&    A_\mu(x,y) = \sum_{i}a_{i}^\dagger(x,y)\p_\mu a_{i}(x,y),  \nonum
&&     \Phi(x,y) = \sum_{i}a_{i}^\dagger(x,y)\,(-a_i(x,y)M(y)
            + M(y)a_i(x,-y)), \nonum
&&  G_\mu(x)=\sum_{j}b_{j}^\dagger(x)\p_\mu b_{j}(x).
  \label{2.6}
\ea
$A_\mu(x,y)$, $\Phi(x,y)$ and $G_\mu(x)$ are identified
the gauge field in the flavor symmetry, Higgs fields,
and the color gauge field in the strong interaction,
respectively. As known in Section 3, it should be noted that
$G_\mu(x)$ is built to commute with $A_\mu(x,y)$ and $\Phi(x,y)$
because $b_j(x)$ commutes with both $a_i(x,y)$ and $M(y)$.\par
In order to identify  $A_\mu(x,y)$ and $G^{}_\mu(x)$ as true gauge fields,
the following conditions have to be imposed.
\ba
&&    \sum_{i}a_{i}^\dagger(x,y)a_{i}(x,y)= 1,  \nonum
&&     \sum_{j}b_{j}^\dagger(x)b_{j}(x)={1\over g_s},
  \label{2.7}
\ea
where $g_s$ is a constant related to
the corresponding coupling constant as shown later.
In general, we can put the right hand side of the first equation
in Eq.(\ref{2.7})
to be ${1/g_y}$. However, we put it as it is to avoid the complexity.\par
Before constructing the gauge covariant field strength,
we address the gauge transformation
of $a_i(x,y)$ and $b_j(x)$ which is defined as
\ba
&&      a^{g}_{i}(x,y)= a_{i}(x,y)g_f(x,y), \nonum
&&      b^{g}_{j}(x) =  b_j(x)g_c(x),
\label{2.8}
\ea
where
$g_f(x,y)$ and $g_c(x)$ are the gauge functions
with respect to the corresponding
flavor unitary group and the color SU(3)$_c$ group, respectively.
It should be noticed that $g_c(x)$ can be taken to commute with $a_i(x,y)$ and
$M(y)$ and at the same time $g_f(x,y)$ is taken to commute with $b_j(x)$.
$g_f(x,y)$ and $g_c(x)$ commute with each other.
Then, we can get the gauge transformation of ${\cal A}(x,y)$ to be
\ba
{\cal A}^g(x,y)\a=g^{-1}_f(x,y) g_c^{-1}(x){\cal A}(x,y)g_f(x,y)g_c(x)\nonum
\a \hskip 2cm +g^{-1}_f(x,y){\bf d}g_f(x,y)+ {1\over g_s}
\,g^{-1}_c(x){d}g_c(x), \label{2.9}
\ea
where use has been made of Eq.(\ref{2.1}) and Equation (\ref{2.8}),
and  as in Eq.(\ref{2.2}),
\ba
       {\bf d}g_f(x,y)\a=
      \partial_\mu g_f(x,y)dx^\mu
                + [-g_f(x,y)M(y) + M(y)g_(x,-y)]\chi \nonum
  \a=\p_\mu g_f(x,y)dx^\mu+\p_y g_f(x,y)\chi.
        \label{2.10}
\ea
Using Equations (\ref{2.6})and (\ref{2.8}),
we can find the gauge transformations of gauge and Higgs fields as
\ba
\a       A_\mu^g(x,y)=g^{-1}_f(x,y)A_\mu(x,y)g_f(x,y)+
                           g^{-1}_f(x,y)\p_\mu g_f(x,y),  \label{2.11}\\
\a       \Phi^g(x,y)=g^{-1}_f(x,y)\Phi(x,y)g_f(x,-y)+
                             g^{-1}_f(x,y)\p_y g_f(x,y), \label{2.12}\\
\a      G_\mu^g(x)=g^{-1}_c(x)G_\mu(x)g_c(x)+
                           \frac1{g_s}g^{-1}_c(x)\p_\mu g_c(x).  \label{2.13}
\ea
Equation(\ref{2.12}) is very similar to other two equations and so
it strongly indicates that the Higgs field is a kind of gauge field
on the discrete space $M_4\times Z_2$. From Eqs.(\ref{2.10}) and
(\ref{2.12}) it is rewritten as
\be
       \Phi^g(x,y)+M(y)=g^{-1}_f(x,y)(\Phi(x,y)+M(y))g_f(x,-y),
                              \label{2.14}\\
\ee
which makes obvious that
\be
H(x,y)=\Phi(x,y)+M(y) \label{2.15}
\ee
is un-shifted Higgs field whereas $\Phi(x,y)$ denotes shifted one with
the vanishing vacuum expectation value.\par
In addition to the algebraic rules in Equation(\ref{2.2}) we add
one more important rule that
\be
              d_\chi M(y)=M(y)M(-y)\chi            \label{2.16}
\ee
which yields together with Eq.(\ref{2.2}) the nilpotency\footnote{
In calculating the nilpotency,  the following rule is taken into account
that whenever the $ d_\chi$ operation jumps over $M(y)$, minus sign
is attached, for example
\ba
     d_\chi \{a(x,y)M(y)b(x,-y)\}\a=(d_\chi a(x,y))M(y)b(x,-y)
                         + a(x,y)(d_\chi M(y))b(x,-y) \nonum
                         \a \hskip 6cm
                         -a(x,y)M(y)(d_\chi b(x,-y)) \nonumber
\ea
}
of the generalized exterior derivative $\bf d$;
\be
          {\bf d}^2 f(x,y)=(d^2+d_\chi^2)f(x,y)=0.  \label{2.17}
\ee
With these considerations we can construct the gauge covariant field
strength.
\be
  {\cal F}(x,y)= F(x,y) +  {\cal G}(x),
\label{2.18}
\ee
where $F(x,y)$ and ${\cal G}(x)$ are the field strengths
of flavor and color gauge fields, respectively and given as
\ba
&&     F(x,y) = {\bf d}A(x,y)+A(x,y)\wedge A(x,y),     \nonum
&&     {\cal G}(x)   =d\,G(x)+g_s G(x)\wedge G(x),
\label{2.19}
\ea
where it should be noted that
${\bf d}A(x,y)=\sum_i{\bf d}a_i^\dagger(x,y)\wedge {\bf d}a_i(x,y)$
and $d\,G(x)=\sum_j{ d}b_j^\dagger(x)\wedge {d}b_j(x)$
are followed because of the nilpotency of $\bf d$ and $d$.
We can easily find the gauge
transformation of ${\cal F}(x,y)$ as
\be
         {\cal F}^g(x,y)=g^{-1}(x,y){\cal F}(x,y)g(x,y),  \label{2.20}
\ee
where $g(x,y)=g_f(x,y)g_c(x)$.
The algebraic rules defined in Equations (\ref{2.2}), (\ref{2.4})
and (\ref{2.7}) yield
\ba
 F(x,y) &=& { 1 \over 2}F_{\mu\nu}(x,y)dx^\mu \wedge dx^\nu  \nonum
           &&\hskip 1.5cm   + D_\mu \Phi(x,y)dx^\mu \wedge \chi
               + V(x,y)\chi \wedge \chi,
                \label{2.21}
\ea
where
\ba
&&  F_{\mu\nu}(x,y)=\p_\mu A_\nu (x,y) - \p_\nu A_\mu (x,y)
               +[A_\mu(x,y), A_\mu(x,y)],\nonum
&&  D_\mu \Phi(x,y)=\p_\mu \Phi(x,y)  + A_\mu(x,y)(M(y) + \Phi(x,y))\nonum
&&     \hskip 6.5cm            -(\Phi(x,y)+M(y))A_\mu(x,-y),\nonum
&&  V(x,y)= (\Phi(x,y) + M(y))(\Phi(x,-y) + M(-y)) - Y(x,y). \label{2.22}
\ea
$Y(x,y)$ in Eq.(\ref{2.18}) is auxiliary field and expressed as
\be
  Y(x,y)= \sum_{i}a_{i}^\dagger(x,y)M(y)M(-y)a_{i}(x,y),
 \label{2.23}
\ee
which may be independent or dependent of $\Phi(x,y)$
and/or may be a constant field.
In contrast to $F(x,y)$, ${\cal G}(x)$ is simply denoted as
\ba
 {\cal G}(x)&=&{1\over 2}{G}_{\mu\nu}(x)dx^\mu\wedge dx^\nu \nonum
        &=&{1\over 2}\{\partial_\mu G^{}_\nu(x)-\partial_\nu G^{}_\mu(x)
        + g_s[G^{}_\mu(x), G^{}_\mu(x)]\}dx^\mu\wedge dx^\nu.
\label{2.24}
\ea
\par
With the same metric structure on the discrete space
$M_4\times Z_{\mathop{}_{2}}$ as in Ref.\cite{MO1} that
\ba
\a <dx^\mu, dx^\nu>=g^{\mu\nu},\quad
g^{\mu\nu}={\rm diag}(1,-1,-1,-1),\nonum
\a <\chi, dx^\mu>=0,\nonum
\a <\chi, \chi>=-\alpha^2
\label{2.25}
\ea
 we can obtain the gauge invariant
\ymh(YMH)
\ba
{\cal L}_{{\mathop{}_{YMH}}}(x)&=&-{\rm Tr}\sum_{y=\pm}{1 \over g_{y}^2}
< {\cal F}(x,y),  {\cal F}(x,y)>\nonum
&=&-{\rm Tr}\sum_{y=\pm}{1\over 2g^2_y}
F_{\mu\nu}^{\dag}(x,y)F^{\mu\nu}(x,y)\nonum
&&+{\rm Tr}\sum_{y=\pm}{\alpha^2\over g_{y}^2}
    (D_\mu \Phi(x,y))^{\dag}D^\mu \Phi(x,y)  \nonum
&& -{\rm Tr}\sum_{y=\pm}{\alpha^4\over g_{y}^2}
        V^{\dag}(x,y)V(x,y)  \nonum
&&-{\rm Tr}\sum_{y=\pm}{1\over 2g_{y}^2}{ G}_{\mu\nu}^{\dag}(x)
{ G}^{\mu\nu}(x),
\label{2.26}
\ea
where $g_y$ is a constant relating
to the coupling constant of the flavor gauge field and
Tr denotes the trace over internal symmetry matrices including the color,
flavor symmetries and generation space.
The third term in the right hand side is the potential term of Higgs particle.
\par
Here, the remark invoked by Sitarz \cite{Sita} should be ordered about
the term to be able to join Equation(\ref{2.26}). He defined
the new metric $g_{\alpha\beta}$ with $\alpha$ and $\beta$
running over $0,1,2,3,4$ by $g^{\alpha\beta}={\rm diag}(+,-,-,-,-)$.
The fifth index represents the discrete space $Z_2$. Then,
$dx^\alpha=(dx^0,dx^1,dx^2,dx^3,\chi)$ is followed.
The generalized field strength $ F(x,y)$ in Equation(\ref{2.21})
is written by $F(x,y)=F_{\alpha\beta}(x,y)dx^\alpha\wedge dx^\beta$ where
$F_{\alpha\beta}(x,y)$ is denoted in Equation(\ref{2.22}). Then, it is
easily derived that ${\rm Tr}\{g^{\alpha\beta}F_{\alpha\beta}(x,y)\}$
is gauge invariant. Thus, the term
\be
|{\rm Tr}\{g^{\alpha\beta}F_{\alpha\beta}(x,y)\}|^2
=\{{\rm Tr}V(x,y)\}^\dagger\{{\rm Tr} V(x,y)\}    \label{Add3}
\ee
can be added to Eq.(\ref{2.26}). Some comments will be ordered in Section 3.
\par
Let us turn to the fermion sector to construct the Dirac Lagrangian.
We start to define the covariant derivative acting
on the spinor field $\psi(x,y)$ which is the representation
of semi simple group of the corresponding flavor gauge group and SU(3)$_c$.
\be
{\cal D}\psi(x,y)=({\bf d}+ {A}^f(x,y))\psi(x,y),
\label{2.27}
\ee
which we call the covariant spinor one-form and $A^f(x,y)$ is chosen
to make ${\cal D}\psi(x,y)$ gauge covariant.
As known in Ssection$\,3$, we specify $A^f_\mu(x,y)$
to be the differential representation
for $\psi(x,y)$.
 Since the role of $d_\chi$ makes the shift  $\Phi(x,y)
\rightarrow \Phi(x,y)+M(y)$ as shown previously, we define also for
fermion field
\be
d_\chi \psi(x,y)=M(y)\chi\psi(x,y)=M(y)\psi(x,-y)\chi \label{2.28}
\ee
which leads Equation(\ref{2.27}) to
\be
{\cal D}\psi(x,y)=(\p_\mu + A_\mu(x,y)+g_sG_\mu(x))\psi(x,y)dx^\mu
        +H(x,y)\psi(x,-y)\chi.
\label{2.29}
\ee
In deriving Equation(\ref{2.29}), Equation(\ref{2.4}) is used and
$g_s$ is necessary for ${\cal D}\psi(x,y)$ to be gauge covariant.
In this context,
$A^f_\mu(x,y)=A_\mu(x,y)+g_sG_\mu(x)$ is Lee algebra
for the fermion representation of gauge group under consideration.
As $\psi(x,y)$ is subjected to the gauge transformation
\be
       \psi^g(x,y)=g^{-1}(x,y)\psi(x,y),  \label{2.30}
\ee
${\cal D}\psi(x,y)$ becomes gauge covariant thanks to Equations(\ref{2.11}),
(\ref{2.12}), (\ref{2.13}) and (\ref{2.29}).
\be
      {\cal D}\psi^g(x,y) =g^{-1}(x,y){\cal D}\psi(x,y).  \label{2.31}
\ee
\par
In addition, ${\bf d}+ A^f(x,y)$ is Lorentz invariant, and so
${\cal D}\psi(x,y)$ is transformed as a spinor just like $\psi(x,y)$ against
Lorentz transformation.
\par
In order to obtain the Dirac Lagrangian for fermion sector,
the associated spinor one-form is introduced as the counter-part of
Equation(\ref{2.27}) by
\be
{\tilde {\cal D}}\psi(x,y)= \gamma_\mu \psi(x,y)dx^\mu
              +i{\cal O}^\dagger(y)c\ma{Y}^\dagger(y)\psi(x,y)\chi,
\label{2.32}
\ee
where $c_{ \mathop{}_{Y}}(y)$
is a matrix assumed to be dimensionless constant related to the
Yukawa coupling constant between Higgs field and fermions
and $c_{ \mathop{}_{Y}}(y)^\dagger=c_{ \mathop{}_{Y}}(-y)$ is satisfied.
The operator ${\cal O}(y)A$ represents the right operation by $A$ for $y=+$
and the left operation by $A$ for $y=-$, and satisfies
$({\cal O}^\dagger(y)A^\dagger)^\dagger={\cal O}(y)A$,  so that
\be
       \{{\cal O}(+)A\}B=BA,\hskip 2cm
       \{{\cal O}(-)A\}B=AB,  \label{add1}
\ee
where $A$ and $B$ are appropriate square matrices. The introduction of
the operator ${\cal O}(y)$ is needed to insure the Hermiticity of
the Yukawa coupling between Higgs and fermion fields.
With the same inner
products for spinor one-forms as in Ref.\cite{MO1} that
\ba
\a <A(x,y)dx^\mu, B(x,y)dx^\nu>=\bar{A}(x,y)B(x,y)g^{\mu\nu},\nonum
\a <A(x,y)\chi, B(x,y)\chi>=-\bar{A}(x,y)B(x,y)\alpha^2,
\label{2.33}
\ea
with vanishing other inner products
we can get the Dirac Lagrangian.
\ba
{\cal L}\ma{D}(x,y)
     =\a i{\rm Tr}<{\tilde {\cal D}}\psi(x,y),{\cal D}\psi(x,y)>\nonum
=\a i\,{\rm Tr}\,[\,{\bar\psi}(x,y)\gamma^\mu(\p_\mu
+A_\mu(x,y)+g_sG_\mu(x))\psi(x,y) \nonum
\a-\alpha^2{\bar\psi}(x,y)\{{\cal O}(y){c\ma{Y}}(y)\}
                                    H(x,y)\psi(x,-y)\,],
\label{2.34}
\ea
where
Tr is also the trace over internal symmetry matrices including the color,
flavor symmetries and generation space.
The total Dirac Lagrangian is the sum over $y$:
\be
{\cal L}_{\mathop{}_{D}}(x)
      =\sum_{y=\pm}{\cal L}_{\mathop{}_{D}}(x,y),
\label{2.35}
\ee
which is apparently invariant for the Lorentz and gauge transformations.
Equations(\ref{2.26}) and (\ref{2.34}) along with Equation(\ref{2.35})
 are crucially important to reconstruct the
spontaneously broken gauge theory.
\par
With these preparations, we can apply the formalism proposed
in this section to the standard model
and compare it with the Sogami's presentation \cite{Soga}.
\par
\section{ Model Construction}

We first specify the fermion field $\psi(x,y)$ in Eq.(\ref{2.27})
with the existing leptons and quarks
and then decide the generalized gauge field ${A}(x,y)$
in order to give the correct Dirac Lagrangian for the fermion sector
in the standard model. Hereafter, the argument $x$ is often abbreviated
if there is no confusion.
\be
       \psi(x,+)=\left(\matrix{
                                u^r\ma{L}\cr
                                u^g\ma{^{L}}\cr
                                u^b\ma{L}\cr
                                \nu\ma{L}\cr
                                d^r\ma{L}\cr
                                d^g\ma{^{L}}\cr
                                d^b\ma{L}\cr
                                e\ma{L}\cr }
                            \right),
\hskip 1.5cm
       \psi(x,-)=\left(\matrix{
                                u^r\ma{R}\cr
                                u^g\ma{^{R}}\cr
                                u^b\ma{R}\cr
                                 0       \cr
                                d^r\ma{R}\cr
                                d^g\ma{^{R}}\cr
                                d^b\ma{^{R}}\cr
                                e\ma{R}\cr }
                            \right), \label{3.36}
\ee
where subscripts $L$ and $R$ denote the left-handed and
right-handed fermions, respectively and
superscripts $r$, $g$ and $b$ represent the color indices.
It should be noticed that $\psi(x,y)$ has the index for the three generation
and so do the explicit expressions for fermions in the right hand sides
of Equation(\ref{3.36}).  In the strict expressions,
 $u,$ $d,$  $\nu$ and $e$
 in  Eq.(\ref{3.36}) should be replaced by
\be
       u \to \left(\matrix{ u \cr
                            c \cr
                            t\cr}
                            \right),
                           \hskip 1cm
       d \to \left(\matrix{ d \cr
                            s \cr
                            b \cr}
                            \right), \hskip 1cm
       e \to \left(\matrix{ e  \cr
                            \mu \cr
                            \tau\cr}
                            \right), \hskip 1cm
       \nu \to \left(\matrix{ \nu_e  \cr
                              \nu_\mu \cr
                              \nu_\tau\cr}
                            \right),
                            \label{3.37}
\ee
respectively. Thus, $\psi(x,\pm)$ is a vector in the 24-dimensional space.
\par
In order to obtain the Dirac Lagrangian for fermion fields in Eq.(\ref{2.34})
we specify   ${A}_\mu(x,y),$ $\Phi(x,y)$ and $G_\mu(x)$
in Eq.(\ref{2.5}) in the following.
\be
      A_\mu(x,+)=-\frac i2\left\{\sum_{k=1}^3\tau^k\otimes1^4 {A\ma{L}^k}_\mu
                          +\tau^0\otimes aB_\mu\right\}\otimes 1^3,
            \label{3.38}
\ee
where
${A\ma{L}^k}_\mu$ and $B_\mu$ are $SU(2)\ma{L}$
and U(1) gauge fields, respectively
and so $\tau^k$ is the Pauli matrices and $\tau^0$ is $2\times 2$ unit matrix.
$1^3$ represents the unit matrix in the generation space and
$a$ is the U(1) hypercharge matrix
corresponding to $\psi(x,+)$ in Equation(\ref{3.36}) and expressed as
\be
     a=\left(\matrix{ \frac13 & 0 & 0 & 0\cr
                      0 & \frac13 & 0 & 0\cr
                      0 & 0 &    \frac13 &0\cr
                      0 & 0 & 0 & -1 \cr}
                      \right).     \label{3.39}
\ee
\be
      A_\mu(x,-)=-\frac i2
                          bB_\mu\otimes 1^3,
            \label{3.40}
\ee
where
$b$ is the U(1) hypercharge matrix
corresponding to $\psi(x,-)$ in Eq.(\ref{3.36}) and so it is
$8\times8$ diagonal matrix
expressed in
\be
     b={\rm diag}\;(\frac43,\frac43,\frac43,0,-\frac23,-\frac23,-\frac23,-2).
     \label{3.41}
\ee
$G_\mu(x)$ is denoted by
\be
       G_\mu(x)=-\frac i2\sum_{a=1}^8 \tau^0\otimes
             \lambda'^a G_\mu^a\otimes 1^3,  \label{3.42}
\ee
where $\lambda'^a$ is $4\times 4$ matrix made of the Gell-Mann
matrix $\lambda^a$ by adding $0$ components
to fourth line and column.
\be
   \lambda'^a  =\left(\matrix{  & &  & 0\cr
                                & \lambda^a &  & 0\cr
                                &  &   &0\cr
                              0 & 0 & 0 & 0 \cr}
                      \right).     \label{3.43}
\ee
Higgs field $\Phi(x,y)$ is represented in $24\times24$ matrix by
\be
    \Phi(+)=\left(\matrix{ \phi^\ast_0 & \phi^+ \cr
                           -\phi^- &   \phi_0  \cr } \right)
                           \otimes 1^4\otimes 1^3,
                           \hskip 2cm
    \Phi(-)=\left(\matrix{ \phi_0 & -\phi^+ \cr
                           \phi^- &   \phi_0^\ast  \cr } \right)
                           \otimes 1^4 \otimes 1^3.
                           \label{3.44}
\ee
Corresponding to Equation(\ref{3.44}), symmetry breaking function
$M(y)$ is given by
\be
    M(+)=\left(\matrix{   \mu & 0 \cr
                           0 &  \mu  \cr } \right)
                           \otimes 1^4\otimes 1^3,
\hskip 2cm M(-)=M(+)^\dagger.
                           \label{3.45}
\ee
With these specifications, all quantities needed
to give the explicit expression to  ${\cal F}(x,y)$
in Equation(\ref{2.16}) can be explicitly written down as follows.
\ba
&&  {F}_{\mu\nu}(x,+)=-\frac i2\left\{\sum_{k=1}^3\tau^k\otimes 1^4
       {F\ma{L}^k}_{\mu\nu} +\tau^0\otimes a B_{\mu\nu}\right\}\otimes1^3,
         \label{3.46} \\
&&  {F}_{\mu\nu}(x,-)=-\frac i2 b B_{\mu\nu}\otimes1^3, \label{3.47} \\
&&  {G}_{\mu\nu}(x)=-\frac i2\sum_{a=1}^8 \tau\otimes
              \lambda'^aG^a_{\mu\nu}\otimes 1^3, \label{3.48}
\ea
where
\ba
 \a     {F\ma{L}^k}_{\mu\nu}=
    \p_\mu {A\ma{L}^k}_\nu-\p_\nu {A\ma{L}^k}_\mu
    +\epsilon^{klm} {A\ma{L}^l}_\mu {A\ma{L}^m}_\nu, \label{3.49} \\
\a     B_{\mu\nu}=\p_\mu B_\nu-\p_\nu B_\mu,        \label{3.50} \\
\a   G_{\mu\nu}^a=\p_\mu G_\nu^a-\p_\nu G_\mu^a
                  +g_sf^{abc}G_\mu^bG_\nu^c. \label{3.51}
\ea

$D_\mu\Phi(x,y)$ in Equation(\ref{2.22}) is represented in
\be
       D_\mu\Phi(+)=(D_\mu\Phi(-))^\dagger=
       \left\{\p_\mu h' -\frac i2(
       \sum_{k=1}^3\tau^k {A\ma{L}^k}_\mu h'+ h'c B_\mu)\right\}\otimes 1^4
       \otimes1^3,
       \label{3.52}
\ee
where $h'$ and $c$ are given as
\be
        h'=\left(\matrix{ \phi^\ast_0+\mu & \phi^+ \cr
                           -\phi^- &   \phi_0+\mu  \cr } \right),
          \hskip 2cm
        c=\left(\matrix{ -1 & 0 \cr
                          0 & 1  \cr } \right). \label{3.53}
\ee
The matrix $c$ stems from
\be
          \tau^0\otimes a-b=
          \left(\matrix{ -1 & 0 \cr
                          0 & 1  \cr } \right)\otimes1^4
                          =c\otimes 1^4 \label{3.54}
\ee
to insure that Higgs doublet $h=(\phi^+,\phi_0+\mu)^t$ has
plus one  hypercharge  and ${\tilde h}=i\tau^2h^\ast$ minus one.
$V(x,y)$ in Equation(\ref{2.19}) is expressed in
\ba
   \a     V(x,+)=(h'h'^\dagger-\mu^2)\otimes 1^4\otimes1^3, \nonum
   \a     V(x,-)=(h'^\dagger h'-\mu^2)\otimes 1^4\otimes1^3.\label{3.55}
\ea
It should be noticed that $Y(x,y)$ in Eq.(\ref{2.22}) can be estimated
by use of Eq.(\ref{3.45}) to be
\be
    Y(x,\pm)=\sum_ia_i^\dagger(x,\pm)M(\pm)M(\mp)a_i(x,\pm)=\mu^2
            \sum_ia_i^\dagger(x,\pm)a_i(x,\pm)=\mu^2 1^{24}, \label{3.56}
\ee
where use has been made of Equation(\ref{2.7}).
\par
The quartic term of Higgs field remarked by Sitarz \cite{Sita}
is expressed in Equation(\ref{Add3}). In general, this term can be
added to Eq.(\ref{2.26}). However, the minimal Lagrangian
in non-commutative geometry seems to be Eq.(\ref{2.26}), since
Sitarz term is introduced by a different way from that in Eq.(\ref{2.26}).
Taking the inner product of field strength is the standard way to get
the Lagrangian for gauge field. For the time being, we proceed without
the Sitarz term.
Putting above equations into Eq.(\ref{2.26})
and rescaling  gauge and Higgs fields
we can obtain \ymh for the standard model
as follows:
\ba
{\cal L}_{{\mathop{}_{YMH}}}&=&
   -\frac14\sum_{k=1}^3\left(F_{\mu\nu}^k\right)^2
   -\frac14B_{\mu\nu}^2 \nonum
  &&  +|D_\mu h|^2   -\lambda(h^\dagger h-{\mu}^2)^2 \nonum
 && - \frac{1}{4}
      \sum_{a=1}^8{G^a_{\mu\nu}}^{\dagger}{G^a}^{\mu\nu}, \label{3.57}
\ea
where
\ba
   &&   F_{\mu\nu}^k=\p_\mu A_\nu^k-\p_\nu A_\mu^k
         +g\epsilon^{klm}A_\mu^lA_\nu^m,  \label{3.58}\\
   &&   B_{\mu\nu}=\p_\mu B_\nu-\p_\nu B_\mu,\label{3.59}\\
   &&     D^\mu h=[\,\p_\mu-{i\over 2}\,(\sum_k\tau^kg{A^k\ma{L}}_\mu
          + \,\tau^0\,g'B_\mu\,)\,]\,h,
          \hskip 1cm h=\left(\matrix{ \phi^+ \cr
                                      \phi_0+\mu  \cr } \right)
                                             \label{3.60}\\
   &&   G_{\mu\nu}^a=\partial_\mu G_\nu^a-\partial_\nu G_\mu^a
         +g_cf^{abc}G_\mu^bG_\nu^c, \label{3.61}
\ea
with
\ba
  &&     g^2=\frac{g_+^2}{12}, \hskip 1cm
      {g'}^2=\frac{2g_+^2g_-^2}{3g_-^2{\rm Tr} a^2+
      3g_+^2{\rm Tr}b^2 }=\frac{g_+^2g_-^2}{16g_+^2+4g_-^2}, \label{3.62}\\
 &&     \lambda=\frac{g_+^2g_-^2}
         {24(g_+^2+g_-^2)},
         \label{3.63}\\
 &&        g_c^2=g_s^2\frac{g_+^2g_-^2}{6(g_+^2+g_-^2)}. \label{3.64}
\ea
Eq.(\ref{3.62}) yields the Weinberg angle with the parameter
$\delta={g_+}/{g_-}$ to be
\be
          \sin^2\theta\ma{W}=\frac{3}{4(\delta^2+1)}. \label{3.65}
\ee
The gauge transformation affords  the Higgs doublet $h$ to take the form
that
\be
             h=\frac1{\sqrt 2}\left(\matrix{ 0 \cr
                              \eta+v  \cr } \right)  \label{3.66}
\ee
which makes possible along with Equations (\ref{3.57})$\sim$(\ref{3.63})
to expect the gauge boson and Higgs particle masses.
\ba
 &&       m\ma{W}^2=\frac1{48}g_+^2v^2, \label{3.67}\\
 &&       m\ma{Z}^2=\frac{1+\delta^2}{12(4\delta^2+1)}g_+^2v^2, \label{3.68}\\
 &&       m\ma{H}= \frac1{12(\delta^2+1)}g_+^2v^2.           \label{3.69}
\ea
These estimations are only valid in the classical level.
However, we are tempted
to compare them with the experimental values.
At first, in the limit of equal coupling constant $g_+=g_-$,
$\sin^2\theta\ma{W}=3/8$ is followed,
which is the same value as in $SU(5)$ or $SO(10)$
GUT. This fact makes possible to expect that the limit of $g_+=g_-$
or $\delta=1$ yields the relations that hold at the grand unification
scale. At the same limit, Equations(\ref{3.67})$\sim$(\ref{3.69}) become
\be
   m\ma{W}=\frac1{\sqrt {48}}g_+v,\hskip 1cm
   m\ma{Z}=\frac1{\sqrt {30}}g_+v,\hskip 1cm
   m\ma{H}=\frac1{\sqrt {24}}g_+v.   \label{3.70}
\ee
{}From Equation (\ref{3.70}) we can get the mass relation
$m\ma{H}=\sqrt{2}m\ma{W}$ which is expected to be hold at the energy of
GUT scale. In general, by eliminating $\delta$ and $g_+v$ from
Equations (\ref{3.65}), (\ref{3.67}) and (\ref{3.69}) the following
interesting relation is extracted.
\be
       m\ma{H}=\frac4{\sqrt 3}m\ma{W}\sin\theta\ma{W}.  \label{3.71}
\ee
Inserting the experimental values $m\ma{W}=79.9$GeV
and $\sin^2\theta\ma{W}=0.233$ we can obtain the tentative value of
the Higgs mass  $m\ma{H}=89$GeV which is sufficiently low that
it is within the energy range of the accelerator in very near future.
\par
If Sitarz term Eq.(\ref{Add3}) would be effective, Equations (\ref{3.69})
and (\ref{3.71}) would become useless. However, we can consider as follows.
We can get the renormalizable Lagrangian for boson sector with
no restrictions between coupling constants if the Sitarz term is
taken into account. However, as stated above,
we can consider Equation (\ref{3.70}) is
still held at the grand unification scale.
Then, the renormalization group equation can be used
to predict the Higgs mass with the initial condition in Equation (\ref{3.70})
at the energy of GUT scale.
This seems to be interesting theme which will be considered in
the different article.
\par
Let us turn to the construction of the Dirac Lagrangian for fermion sector.
After the rescaling of the boson fields,
corresponding with the specification of Eq.(\ref{3.36})
we can write the covariant spinor
one-form in Eq.(\ref{2.29}) and the associated spinor one-form in
Eq.(\ref{2.32}) as
\ba
\hskip -0.5cm {\cal D}\psi(x,+)\a= \left[ \p_\mu\otimes1^8
 -\frac i2 \{g\sum_{k=1}^3\tau^k\otimes1^4{A^i\ma{L}}_\mu
   +\tau^0\otimes ag'B_\mu)+\right. \nonum
  \a \left.\sum_{a=1}^8\tau^0\otimes\lambda'^ag_c G^a_\mu
   \}\right]\otimes1^3\psi(x,+)dx^\mu
    +g\ma{H} h\otimes 1^4\otimes 1^3 \psi(x,-)\chi
              \label{3.72}\\
\hskip -0.5cm{\tilde {\cal D}}\psi(x,+)\a=
1^{24}\gamma_\mu \psi(x,+)dx^\mu
              +i\{{\cal O}^\dagger(+)c\ma{Y}^\dagger\}\psi(x,+)\chi,
              \label{3.73}
\ea
and
\ba
 {\cal D}\psi(x,-)\a= \left[1^8\p_\mu
 -\frac i2\{ bg'B_\mu\psi(x,-)
                 +\sum_{a=1}^8\tau^0\otimes\lambda'^ag_c G^a_\mu\}\otimes1^3
                 \psi(x,-)dx^\mu\right.\nonum
 \a  \left. + h^\dagger\otimes1^4\otimes1^3\psi(x,+)\chi\right].
     \label{3.74}\\
{\tilde {\cal D}}\psi(x,-)\a=
1^{24}\gamma_\mu \psi(x,-)dx^\mu
              +i\{{\cal O}^\dagger(-)c\ma{Y}\}\psi(x,-)\chi.
\label{3.75}
\ea
According to Equation(\ref{2.34})
we can find with a special attention on Equation(\ref{add1})
the Dirac Lagrangian for the standard model as follows:
\ba
{\cal L}\ma{D}\a=\sum_{y=\pm}
     i<{\tilde {\cal D}}\psi(x,y),{\cal D}\psi(x,y)>\nonum
   \a=i{\bar \psi}(x,+)\gamma^\mu\left[1^8\p_\mu-\frac i2\{g\sum_{k=1}^3
        {A\ma{L}^k}_\mu\otimes1^4+g'\tau^0\otimes aB_\mu\right. \nonum
   \a \hskip 6.5cm \left.+g_c\sum_{a=1}^8\tau^0\otimes\lambda'^aG_\mu\}\right]
                         \otimes 1^3\psi(x,+)\nonum
   \a+i{\bar \psi}(x,-)\gamma^\mu\left[1^8\p_\mu-\frac i2\{
        g' bB_\mu  +g_c\sum_{a=1}^8\tau^0\otimes\lambda'^aG_\mu\}\right]
         \otimes1^3\psi(x,-)\nonum
    \a -{\bar \psi}(x,+) h'\otimes1^4\otimes1^3g\ma{Y}\psi(x,-)-
    {\bar \psi}(x,-)g\ma{Y}^\dagger h'^\dagger\otimes1^4\otimes1^3\psi(x,+),
          \label{3.76}
\ea
which is sufficient as the Dirac Lagrangian of the standard model
with a Yukawa coupling constants
$g\ma{Y}=\alpha^2g\ma{H}c\ma{Y}$ in matrix form given as
\be
    g\ma{Y}={\rm diag}(g^u,g^u,g^u,g^\nu,g^d,g^d,g^d,g^e).
                               \label{3.78}
\ee
$g^u$, $g^d$, $g^\nu$ and $g^e$ in Equation(\ref{3.78}) are
complex Yukawa coupling constant written in
$3\times3$ matrix in generation space.
Equation(\ref{3.78}) yields the interaction Lagrangian between Higgs
and fermion fields as
\be
{\cal L}\ma{Yukawa}=-{\bar q}\ma{L}h\otimes 1^3\otimes g^d\,d\ma{R}
         -{\bar q}\ma{L}{\tilde h}\otimes1^3\otimes g^u\,u\ma{R}
         -{\bar l}\ma{L}h\otimes g^e\, e\ma{R} -h.c., \label{3.79}
\ee
where
\be
             q\ma{L}=\left(\matrix{ u\ma{L} \cr
                                    d\ma{L}  \cr } \right),\hskip 2cm
             l\ma{L}=\left(\matrix{ \nu\ma{L} \cr
                                    e\ma{L}  \cr } \right)
  \label{3.80}
\ee
with color indices abbreviated.
In this notation it should be noted again that in the exact notation
$u$, $d$, $e$ and $\nu$ are replaced by
\be
       u \to \left(\matrix{ u \cr
                            c \cr
                            t\cr}
                            \right),
                           \hskip 1cm
       d \to \left(\matrix{ d \cr
                            s \cr
                            b \cr}
                            \right), \hskip 1cm
       e \to \left(\matrix{ e  \cr
                            \mu \cr
                            \tau\cr}
                            \right), \hskip 1cm
       \nu \to \left(\matrix{ \nu_e  \cr
                              \nu_\mu \cr
                              \nu_\tau\cr}
                            \right).
                            \label{3.81}
\ee
If we diagonalize the complex Yukawa coupling constants
$g^u$ and $g^d$, we can obtain the correct Dirac Lagrangian for
the standard model which contains the generation mixing
through the Kobayashi-Maskawa mixing matrix. $g^\nu$ is taken zero and
$g^e$ is a matrix already diagonalized, which yield that
there is no generation mixing for lepton family.
\section{ Concluding remarks}
We present the new incorporation of the strong interaction in our NCG scheme
to reconstruct the standard model. It begins with the formation of the
fermion field which is extracted from that of SO(10) GUT by discarding the
anti-fermion part. Then, the generalized gauge field including the left-handed,
U(1) and color gauge fields is constructed in order to yield the correct
Dirac Lagrangian for fermion sector in the standard model.
Characteristic points is that all gauge fields and Higgs field are written
in $24\times24$ matrices according to 24-dimensional vector of
the fermion field. The gauge fields are constructed so as to commute
with each other.\par
The generation mixing is taken into account through the associated spinor
one-form in Equation(\ref{2.32}) so that it does not give any effect on
the \ymh. As a result, we can estimate gauge boson and Higgs masses
without any effect of the generation mixing matrices as
in Equation(\ref{3.70}), which is very
characteristic point of our approach compared with other approaches
\cite{Cham}, \cite{KDC}, \cite{Soga}. For example, Sogami \cite{Soga}
estimated the Higgs mass around $m\ma{H}={\sqrt {2}}m_t$ with the top quark
mass $m_t$ since his generalized covariant derivative contains
the complex Yukawa couplings between fermion and Higgs fields.
$m\ma{H}={\sqrt {2}}m_t$ is a characteristic prediction in such approach,
which is contractive to our result
$m\ma{H}=(4/\sqrt 3)m\ma{W}\sin\theta\ma{W}$.
Though these predictions are only valid in tree level, it is expected that
not so much differences seem to appear even in quantum effect included.
Thus, Higgs search experiments in very near future shall judge
which is superior to other on the premise that NCG approach is true
understanding of the Higgs mechanism.\par
If we include in the \ymh the quartic term of Higgs field due to Sitarz
\cite{Sita},
\ymh is free from any restriction for coupling constants.
Then, we can reproduce the renormalizable Lagrangian for boson sector
of the standard model on our NCG approach. In the limit of
$g_+=g_-$, we can get the prediction of $\sin^2\theta\ma{W}=3/8$, which
suggests that our \ymh in this limit and with the vanishing Sitarz'term
is the Lagrangian at the energy of GUT scale.
The Higgs mass $m\ma{H}={\sqrt 2}m\ma{W}$ is followed in this limit.
Thus, it is expected to calculate the Higgs mass by use of
the renormalization equation in the same way as the Weinberg angle
was calculated. This calculation will be performed in very near future.
\vskip 0.5 cm
\begin{center}
{\bf Acknowledgement}
\end{center}
The author would like to
express their sincere thanks to
Professors J.~IIzuka,
 H.~Kase, K. Morita and M.~Tanaka
for useful suggestion and
invaluable discussions on the non-commutative geometry.
\begin{appendix}
\section{Appendix}
In this Appendix, we explain rather un-known property of
the extra-differential $\chi$ which is introduced with respect to
the discrete space $Z_2$ in more extended version than that necessary in
this article.
$\chi$ plays a role that connects the fields
on  $y=+$ and $y=-$ spaces and so exhibits  the following equation.
\be
         f(x,y)\chi=\chi f(x,-y), \label{A1}
\ee
where $f(x,y)$ can be gauge field, Higgs field or fermion field.
The generalized gauge fields $A(x,y)$ is expressed as
\be
      A(x,y)=A_\mu(x,y)dx^\mu+\Phi(x,y)\chi. \label{A2}
\ee
Let $a_i(x,y)$ and $a_i(x,-y)$ in Eq.(\ref{2.3}) be in general
the square matrices
with $(p,p)$ and $(q,q)$ types, respectively.
Then, $M(y)$ is the $(p,q)$ type matrix to lead to
the consistent calculation  in Eq.(\ref{2.1}).
To be general, $p\ne q$ is required to reconstruct the realistic gauge
theory with the spontaneous symmetry breaking such as the standard model
or SU(5) GUT. For example, the symmetry of the standard model is
SU$(2)_{\mathop{}_{L}}\times$U$(1)_{\mathop{}_{Y}}$, so that
$p=2$ and $q=1$ should be followed in our previous formulation\cite{MO1},
\cite{OSM}.
Thus,  $\Phi(x,y)$ is in general not square matrix.
The reader might imagine why $\Phi(x,y)$ is not square matrix
in spite that $A_\mu(x,y)$ is square matrix in Eq.(\ref{A2}).
This is because there is no linear transformation between
$dx^\mu$ and $\chi$  and so the components of
$A_\mu(x,y)$ and $\Phi(x,y)$ never mix with each others. Our algebra
in NCG on the discrete space is consistently constructed in conformity with
this fact. \par
Furthermore, to justify Eq.(\ref{A1}),
We take the example of $A(x,y)\wedge A(x,y)$.
Eq.(\ref{A1})
leads this wedge product to
\ba
   A(x,y)\wedge A(x,y)=\a(A_\mu(x,y)dx^\mu+\Phi(x,y)\chi)\wedge
                        (A_\mu(x,y)dx^\mu+\Phi(x,y)\chi) \nonum
                =\a A_\mu(x,y)A_\nu(x,y)dx^\mu\wedge dx^\nu\nonum
                \a+ (A_\mu(x,y)\Phi(x,y)-\Phi(x,y)A_\mu(x,-y))dx^\mu\wedge\chi
                \nonum
                \a +\Phi(x,y)\Phi(x,-y)\chi\wedge\chi. \label{A3}
\ea
We can readily see the consistency of the matrix products
in Equation (\ref{A3}) for $y=+$ and $y=-$ in the case
of the standard model stated just above.
\par
Eq.(\ref{A1}) was originally introduced by Sitarz \cite{Sita} without
the introduction of the symmetry breaking function $M(y)$. Ref.\cite{Gou}
reconstructed the Weinberg-Salam theory according to Sitarz's formalism
where all fields such as $A_\mu(x,\pm)$ and $\Phi(x,\pm)$ are same type
square matrices. Recently, Konisi and Saito \cite{KS} indicated
that the Higgs field $\Phi(x,y)$ might become the unitary matrix
to lead the vanishing Higgs potential. However, their conclusion
can not be applicable to our case because $A(x,y)$
in our formalism expresses the different kind of gauge field for each
sheet $y=\pm$ and $\Phi(x,y)$ is not in general the square matrix.\par
We can formulate our framework in slightly different way though it yields
the same results as in this article, which may help one understand
the true meaning of Equation(\ref{A1}).
If we express the generalized gauge field ${\cal A}$ without the variable
$y$ in the following matrix form
\be
   {\cal A}=\left(\matrix{ A_\mu(x,+)dx^\mu &  \Phi(x,+)\chi\cr
                       \Phi(x,-)\chi    &  A_\mu(x,-)dx^\mu\cr}
                                   \right), \label{A4}
\ee
we can calculate the wedge product of ${\cal A}\wedge {\cal A}$ as in
\ba
 \hskip -1cm
 {\cal A}\a\wedge {\cal A}= \nonum
 \a \nonum
 \a \hskip -1cm
  \left(\matrix{  \hskip -1.5cm A_\mu(x,+)A_\nu(x,+)dx^\mu\wedge dx^\nu
                          & \hskip -3.5cm (A_\mu(x,+)\Phi(x,+) \cr
                    \hskip 2cm  +\Phi(x,+)\Phi(x,-)\chi\wedge\chi &
             \hskip 2cm    -\Phi(x,+)A_\mu(x,-))dx^\mu\wedge\chi \cr
            \hskip -2.5cm (A_\mu(x,-)\Phi(x,-) &
            \hskip -1cm  A_\mu(x,-)A_\nu(x,-)dx^\mu\wedge dx^\nu \cr
          \hskip 2cm   -\Phi(x,-)A_\mu(x,+))dx^\mu\wedge\chi &
          \hskip 2cm   +\Phi(x,-)\Phi(x,+)\chi\wedge\chi \cr
                             } \right) \label{A5}
\ea
with the alternative rule that
\be
         f(x,y)\chi=\chi f(x,y) \label{A6}
\ee
in this case.
Since \ymh ${\cal L}\ma{YMH}$ is defined with the field strength
$\cal F$ expressed in
\be
        {\cal F}={\bf d}{\cal A}+{\cal A}\wedge {\cal A} \label{A8}
\ee
through the equation
\be
          {\cal L}\ma{YMH}={\rm Tr}<{\cal F}, {\cal F}>,  \label{A7}
\ee
we can find the same Lagrangian in Section 3.
Thus, we can conclude that Equation(\ref{A1}) makes possible to
perform the calculations in our formalism without the matrix product
as in Equation(\ref{A5}). Then, it becomes evident that Eq.(\ref{A1})
is never the relation between the matrix elements of $f(x,\pm)$
and it realizes the non-commutativity of our algebra in our formalism.
\end{appendix}

\def\jmp{J.~Math.~Phys.$\,$}
\def\pl{Phys. Lett.$\,$ }
\def\np{Nucl. Phys.$\,$}
\def\ptp{Prog. Theor. Phys.$\,$}
\def\prl{Phys. Rev. Lett.$\,$}
\def\pr{Phys. Rev. D$\,$}
\def\mp{Int. Journ. Mod. Phys.$\,$ }

\end{document}